\documentclass[sigconf]{acmart} 
\pdfoutput=1
\usepackage{microtype}
\usepackage{graphicx}
\usepackage{subfigure}
\usepackage{balance}
\usepackage{booktabs} 
\usepackage{amsthm}
\usepackage{array}
\usepackage{graphicx}
\usepackage{clrscode}
\usepackage{subfigure}
\usepackage{multirow}
\usepackage{multicol}
\usepackage{float}
\usepackage{color}
\usepackage{xcolor}
\usepackage{amsopn}
\usepackage{mathrsfs}
\usepackage{mathtools}
\usepackage{amsmath}
\usepackage{arydshln}
\usepackage{hyperref}
\usepackage{blkarray}
\usepackage{enumerate}
\usepackage{courier}
\usepackage{rotating}
\usepackage{bm}
\usepackage{wrapfig}
\usepackage{subfigure}
\usepackage{array}
\usepackage{ragged2e}
\usepackage{hyperref}
\usepackage{marvosym}
\usepackage{threeparttable}
\usepackage{adjustbox} 
\usepackage{enumitem}
 
\usepackage{comment}
\usepackage{newfloat}
\usepackage{listings}

\definecolor{codegreen}{rgb}{0,0.3,0.6}
\definecolor{codegray}{rgb}{0.5,0.5,0.5}

\floatstyle{ruled}
\newfloat{listing}{tb}{lst}{}
\floatname{listing}{Listing}

\usepackage[ruled,linesnumbered]{algorithm2e}

\newcommand{\ie}{\emph{i.e.,}\xspace}
\newcommand{\eg}{\emph{e.g.,}\xspace}

\newcommand{\paratitle}[1]{\vspace{1.5ex}\noindent\textbf{#1}}

\newcommand{\ignore}[1]{}

\SetKwComment{comment}{ $triangleright$ \ }{}

\theoremstyle{definition}

\theoremstyle{theorem}

\theoremstyle{proof}

\theoremstyle{remark}

\AtBeginDocument{%
  \providecommand\BibTeX{{%
    \normalfont B\kern-0.5em{\scshape i\kern-0.25em b}\kern-0.8em\TeX}}}

\setcopyright{acmcopyright}
\copyrightyear{2022}
\acmYear{2022}
\acmDOI{10.1145/1122445.1122456}

\acmConference[Woodstock '21]{Woodstock '18: ACM Symposium on Neural
  Gaze Detection}{June 03--05, 2018}{Woodstock, NY}
\acmBooktitle{Woodstock '21: ACM Symposium on Neural Gaze Detection,
  June 03--05, 2021, Woodstock, NY}
\acmPrice{15.00}
\acmISBN{978-1-4503-XXXX-X/18/06}

\let\emph\textit 

\begin{document}
\title{Recent Advances in RecBole: Extensions with more \\Practical Considerations}
\author{Lanling Xu$^{1,3\dagger}$, Zhen Tian$^{1,3\dagger}$, Gaowei Zhang$^{1,3}$, Lei Wang$^{1,3}$, Junjie Zhang$^{1,3}$,\\ Bowen Zheng$^{3}$, Yifan Li$^{3}$, Yupeng Hou$^{1,3}$, Xingyu Pan$^{2,3}$, Yushuo Chen$^{1,3}$,\\ Wayne Xin Zhao$^{1,3}$\textsuperscript{\Letter}, Xu Chen$^{1,3}$\textsuperscript{\Letter}, and Ji-Rong Wen$^{1,2,3}$}
\thanks{$\dagger$ Both authors contributed equally to this work. }
\thanks{\textsuperscript{\Letter} Wayne Xin Zhao (batmanfly@gmail.com) and Xu Chen (successcx@gmail.com) are the corresponding authors.}

\affiliation{%
  \institution{\{$^1$Gaoling School of Artificial Intelligence, $^2$School of Information\},  Renmin University of China, China}
  \institution{$^3$Beijing Key Laboratory of Big Data Management and Analysis Methods, China}
  \country{}
}

\renewcommand{\shortauthors}{Xu and Tian, et al.}

\begin{abstract}
RecBole has recently attracted increasing attention from the research community.
As the increase of the number of users, we have received a number of suggestions and update requests.
This motivates us to make some significant improvements on our library, so as to meet the user requirements and contribute to the research community. 
In order to show the recent update in  RecBole,  
we write this technical report to introduce our latest improvements  on RecBole.
In general, we focus on the flexibility and efficiency of RecBole in the past few months.
More specifically, we have four development targets: 
(1) more flexible data processing,
(2) more efficient model training,
(3) more reproducible configurations, and
(4) more comprehensive user documentation.
Readers can download the above updates at: \url{https://github.com/RUCAIBox/RecBole}.
\end{abstract}

\keywords{Recommender systems, Open-source library, Benchmarking}
\maketitle

\section{INTRODUCTION}
In recent years, recommender systems have received increasing attention from both academia and industry. 
Despite the great success, the reproducibility has long been a severe concern in the literature~\cite{worry_neural,cremonesi2020methodological}. 
For alleviating this problem, recent years have witnessed a series of open source benchmarking libraries such as DaisyRec~\cite{sun2022daisyrec}, TorchRec~\cite{TorchRec}, EasyRec~\cite{EasyRec} and RecBole~\cite{zhao2021recbole,zhao2022recbole}.

Among these benchmarking libraries, RecBole~\cite{zhao2021recbole} is featured with the unified benchmarking framework, standardized models and datasets, extensive evaluation protocols, efficient training and user-friendly documentation. Since the first release in 2020, RecBole has received nearly 2300 stars and 425 forks on GitHub. Besides, we are committed to solving the regular usage problems by handling more than 400 issues and 900 pull requests. 
During this period, our team has also developed a series of up-to-date algorithms to facilitate latest research, which have been included in RecBole~2.0~\cite{zhao2022recbole}.

\begin{figure}
    \centering
    \includegraphics[width=1.0\linewidth]{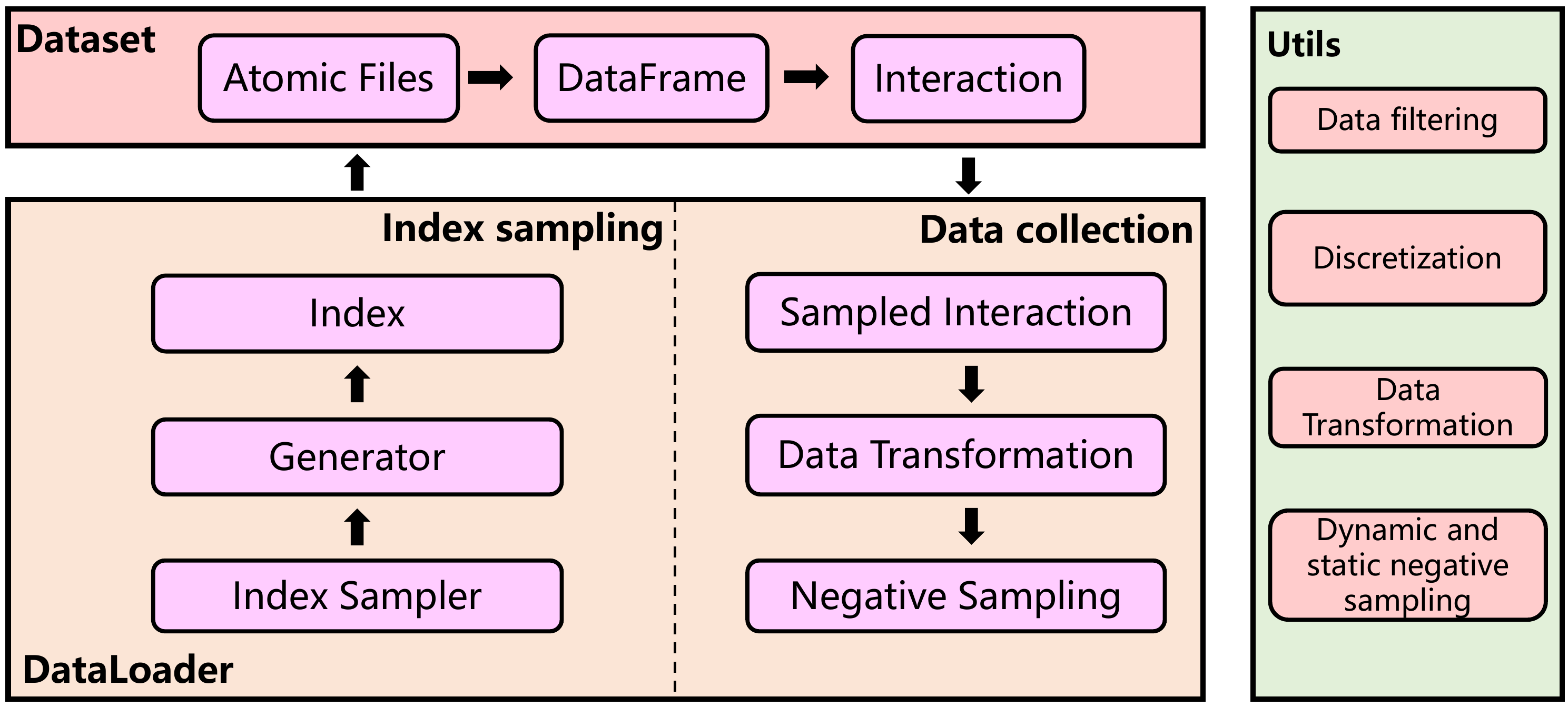}
    \caption{The overall framework of data module in RecBole.}
    \label{figs:dataframework}
\end{figure}

As a user-friendly recommendation library, RecBole not only keeps up with the most recent mainstream progress on recommendation, but also continually optimizes its design for more flexible and easier use. 
For this purpose, we update several commonly used mainstream data processing methods and reframe the data module to be compatible with a series of efficient APIs. Meanwhile, we implement distributed training and parallel tuning modules for model acceleration over large-scale data. Furthermore, We provide more benchmarking datasets, well-built parameter configurations and detailed usage documents, making it easier to use our library.

In order to make RecBole a more user-friendly benchmarking library for recommendation, 
our team has considered the up-to-date requests in version 1.1.1, with the following four highlights:

$\bullet$ \textbf{More flexible data processing}. To satisfy diverse data processing requests, we largely update the data module with a more flexible processing pipeline.
For extensibility, we reframe the overall {data flow} with PyTorch~\cite{paszke2019pytorch}. Considering the varied features of different recommendation tasks, we add data transformation for sequential models, discretization of continuous features for context-aware models and knowledge graph filtering for knowledge-aware models. We also optimize the sampling module to support both static and dynamic negative samplers.
These updated features enable researchers to easily implement the mainstream data processing methods so as to conduct research in a more convenient way.

$\bullet$ \textbf{More efficient training and tuning}. GPU-based acceleration is one of our major features, and we significantly improve the efficiency in this update through three new strategies, \ie multi-GPU training and evaluation, mixed precision training and intelligent hyper-parameter tuning, which make it more efficient to deal with  large-scale interaction data in different recommendation scenarios.

$\bullet$ \textbf{More reproducible configurations}. In the field of recommender system, it is important to establish reproducible benchmarking for performance comparison, while the model results largely depend on the selected dataset and hyper-parameter configurations.
In this update, we add 13 new processed datasets with unified atomic files on the basis of 28 existing ones, which can be directly used as the input of RecBole.
To further facilitate the search process of hyper parameters, we provide the hyper-parameter selection range and recommended configurations for each model on three datasets, covering four types of recommendation tasks. 
Benefiting from the appropriate parameter configurations, researchers can easily reproduce and compare baseline models with our library.  

$\bullet$ \textbf{More user-friendly documentation}. 
To make our library more accessible, we update the websites and documentation with  detailed descriptions.
In particular, we integrate new features of RecBole~2.0~\cite{zhao2022recbole} into the website, and add instructions of the customized training strategy, multi-GPU training cases and detailed running examples in our handbook.
With this updated documentation, new users can get started with our library quickly.

In the following, we first detail each of the above four features in this update. Then we briefly discuss the development goals and future directions of RecBole.

\section{UPDATE HIGHLIGHTS}

\subsection{Flexible Data Processing}

In the previous version of RecBole~\cite{zhao2021recbole}, we have provided a unified data processing pipeline. In this update, we aim to improve this pipeline from three aspects, \ie more compatible data module, customized data transformation and task-oriented data pre-processing.

\subsubsection{Compatible Data Module}
To support multiple types of recommendation tasks in a unified way, we design a more general data flow under the framework of PyTorch, which can easily  implement  mainstream data processing methods.
As shown in Figure~\ref{figs:dataframework}, our data module mainly includes two derived classes: \textsf{Dataset} and \textsf{DataLoader}, which are inherited from \textsf{torch.utils.data.Dataset} and \textsf{torch.utils.data.DataLoader}, respectively.
The data flow of \textsf{Dataset} is described as follows: $Atomic \, Files \rightarrow DataFrame \rightarrow Interaction$, where the formatted dataset files are loaded and processed to derive the internal structure \textsf{Interaction}. 
Specially, during the transformation process from atomic files to \textsf{pandas.DataFrame}, we incorporate a series of intermediate data filtering and feature processing methods to derive the target instances, which can be subsequently transformed to \textsf{Interaction} for iteration and indexing.

The class \textsf{DataLoader} provides an iterable method over the given class \textsf{Dataset}.
For compatibility, we divide the process of each iteration into two stages: \emph{index sampling} and \emph{data collection}. 
The former generates the sampled index of \textsf{Interaction}, which is mainly implemented by the API provided by the base class of \textsf{DataLoader} in PyTorch.
The latter is implemented in the interface \textsf{collate\_fn($\cdot$)}, which first retrieves the sampled \textsf{Interaction} based on the sampled index. 
To support various recommendation tasks, we apply a series of flexible data transformation and negative sampling strategies to generate training and testing data for evaluation.

\subsubsection{Customized Data Transformation} In this  update, we add a new variable ``\textsf{transform}'' in the dataloader to perform useful transformations for sequential models, which are independent of the model implementation.
Specifically, we add four major augmentation operations for the processing procedure of data~\cite{2020Contrastive}.

$\bullet$~\textsf{MaskItemSequence} is 
the token masking operation proposed in NLP~\cite{BERT}, which has also been used in recommender system, such as BERT4Rec~\cite{bert4rec} and S$^3$Rec~\cite{S3Rec}. Here, we introduce this transformation method for interaction sequences as in \cite{S3Rec,bert4rec}. For each user's historical sequence, this method replaces a proportion of items by a special symbol \textsf{[mask]} to get a masked sequence.

$\bullet$~\textsf{FlexiblePadLocation} provides an option to specify the position of \textsf{[pad]} to align the sequence length, and users can choose to place it at the beginning or end of a sequence. 

$\bullet$~\textsf{CropItemSequence}~\cite{2020Contrastive} generates  local views on historical interaction sequences by selecting a continuous sub-sequence.

$\bullet$  \textsf{ReorderItemSequence} alters the order of items in the user sequence by randomly shuffling their positions~\cite{2020Contrastive}. 

Besides, as a user-defined interface, \textsf{UserDefinedTransform} is reserved to facilitate further development by the users themselves.

\subsubsection{Task-oriented Data Pre-processing} 
To be more compatible with recent research in recommendation,
we update the task-oriented interfaces for knowledge graph in knowledge-aware models and numerical features in context-aware models. Meanwhile, we implement both static and dynamic samplers for sample-based methods.

$\bullet$ \textbf{Filtering and augmentation for knowledge graph}.
Since the knowledge graph is very large, we add a new mechanism to support the filtering process of entities and relations in the triples.
As shown in~\cite{wang2019kgat, wang2021kgin}, inverse relations are often used in knowledge-aware models,  we also provide a way for adding inverse relations.
Considering the two requirements, we implement two specific methods for pre-processing knowledge graphs: (1) $k$-core filtering is performed on the nodes and relations to mitigate the noise problems, and (2)  users can choose whether to construct the inverse relations for triplets as data augmentation. 

$\bullet$ \textbf{Discretization of continuous features}. For numerical features, we support two mainstream representation approaches (\ie  field embedding and discretization) in the latest version of RecBole,
which are widely used in existing works~\cite{guo2021embedding}.
Specially, field embedding is a commonly used approach in academia, where features in the same field share a uniform embedding and multiply it with their scale values.
Recent industrial recommender system tends to utilize discretization methods to handle numerical features.
By discretization, numerical features are transformed to different discrete value buckets.
We implement two widely used discretization methods, \ie \textit{equal distance discretization} and \textit{logarithm discretization}.

\textit{Equal distance discretization} divides the numerical features into multiple buckets with the same width. 
The formal representation of equal distance discretization could be defined as:
\begin{equation}
    \hat{x}_i = {\rm{floor}} \bigg(\frac{x_i - x_i^{min}}{x_i^{max} - x_i^{min}} \cdot N_i \bigg),
\end{equation}
where $[x_i^{max} , x_i^{min}]$ denotes the range of the $i$-th numerical field, $N_i$ denotes the number of discretization buckets 
of the $i$-th numerical field, and {$\rm{floor}(\cdot)$ is the rounding down operation.}

{\textit{Logarithm discretization} is proposed by the champion of Criteo advertiser prediction in Kaggle~\cite{guo2021embedding}.} By logarithm discretization, each numerical feature is transformed as:
\begin{equation}
    \hat{x}_i = {\rm floor} \big(\log(x_i)^2 \big).
\end{equation}

\begin{figure}[H]
     \centering
     \includegraphics[width=0.8\linewidth]{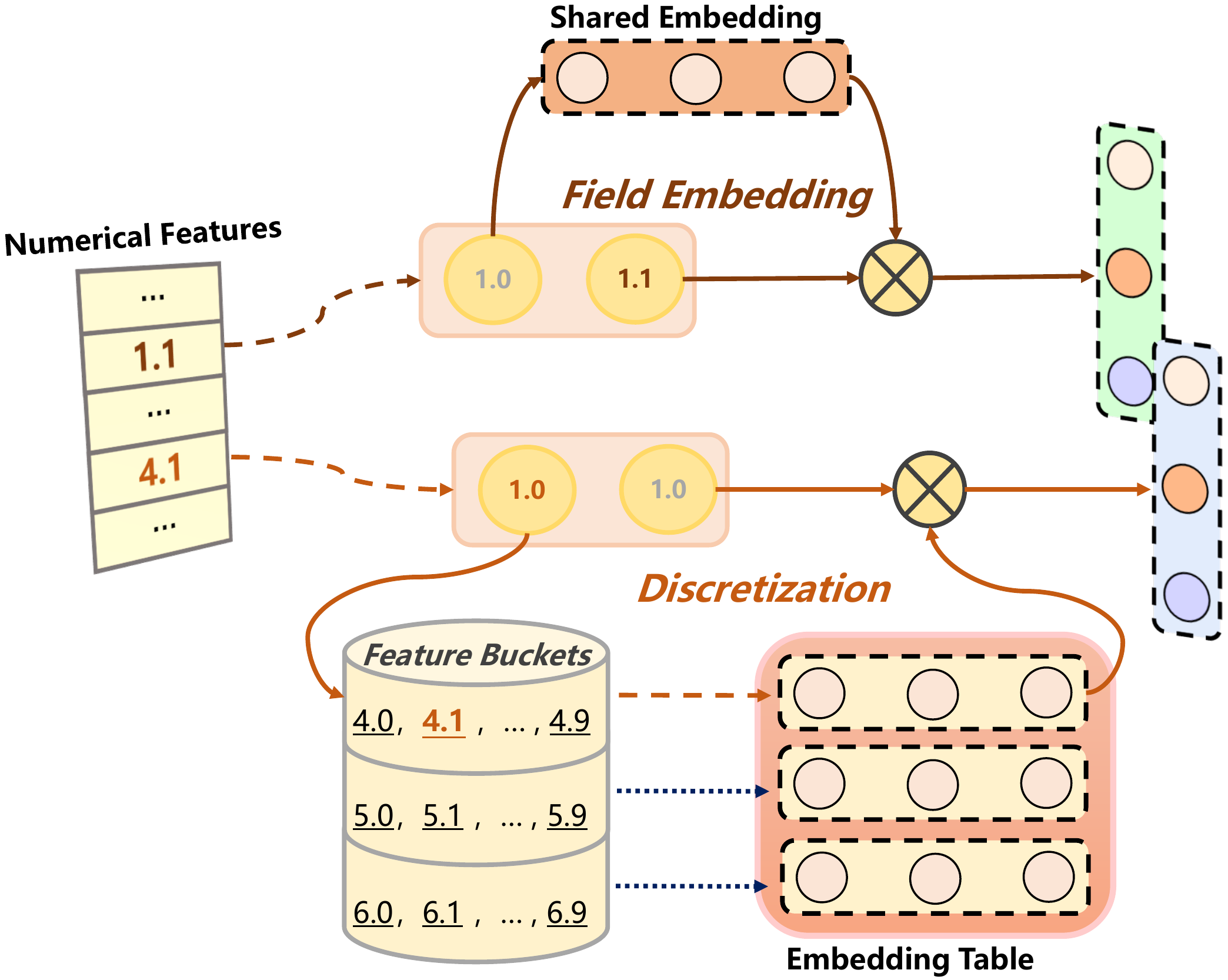}
     \captionsetup{font={small}}
     \caption{The unified embedding protocol for float-like fields.}
     \label{fig:ds}
\end{figure}

To support more representation strategies, we design a standard data structure for numerical features.
As shown in Figure~\ref{fig:ds}, each numerical feature is represented by a tuple consisting of continuous
and discrete values. 
{For the field of embedding, we keep the scale value and set the discrete value to one. While for discretization, the scale value is fixed to one and the discrete value is determined by the discretization method. Therefore, we could utilize a unified embedding protocol to support both field embedding and discretization, \ie features with the same discrete value share the same embedding and multiply it with their continuous values.}

$\bullet$ \textbf{Dynamic and static negative sampling}.
Recent studies show that negative sampling \cite{mao2021simplex}
largely influences the model results. However, existing recommendation libraries either ignore the sample selection strategies, or only consider static negative samplers, which is incapable to support a number of  models based on specific negative sampling methods. 
Considering the above issues, we add dynamic negative sampling method~\cite{zhang2013dns} in our implementation, and support three commonly used  negative sampling  methods,  including random negative sampling~(RNS)~\cite{BPR}, popularity-biased negative sampling~(PNS)~\cite{pns_alpha,PNS} and dynamic negative sampling~(DNS) \cite{zhang2013dns}.
In terms of PNS, we borrow the idea from word vector representation~\cite{word2vec} and set a hyper-parameter $\alpha$ to control the impact of popularity on negative sampling like~\cite{pns_alpha}. Formally, the probability of an item to be  sampled as the negative for a user is $p(i)\propto deg(i)^{\alpha}$, where $deg(i)$ is the total interaction number for  item $i$ by all the users.
By setting different $\alpha$ in ``\textsf{train\_neg\_sample\_args}'', we can implement both low-popularity and high-popularity samplers.

\subsection{Efficient Training and Tuning}

\subsubsection{Efficient GPU utilization.} 

\ignore{In recent years, the design of models has gradually become more and more complex. With the proposal of a series of pre-training models such as BERT~\cite{BERT} and GPT~\cite{GPT}, pre-training large models has gradually become a research hotspot. Although these large models can achieve outstanding performance with a large number of parameters, the large model architecture increases the demand for computing ability of the machine. Obviously, training on a single GPU is no longer able to support the computational requirements of models with \textcolor{red}{the} increasing number of parameters. Therefore, many methods have been proposed to improve the \textcolor{red}{computational efficiency, where distributed computing is one of the most popular one}. 
}

Efficiency is an important factor to consider for a recommendation library. 
In this update, we provide a multi-GPU and multi-machine training method based on \textsf{torch.nn.parallel.DistributedDataParallel}, which  largely improves the efficiency of model training and evaluation. In addition to the normal training and evaluation settings, the user can set up the distributed computing environment~(\eg \textsf{ip}, \textsf{port}, \textsf{world\_size}, \textsf{nproc}, and  \textsf{group\_offset}). We present a usage example in the appendix.

\begin{figure}[H]
    \centering
    \includegraphics[width=0.8\linewidth]{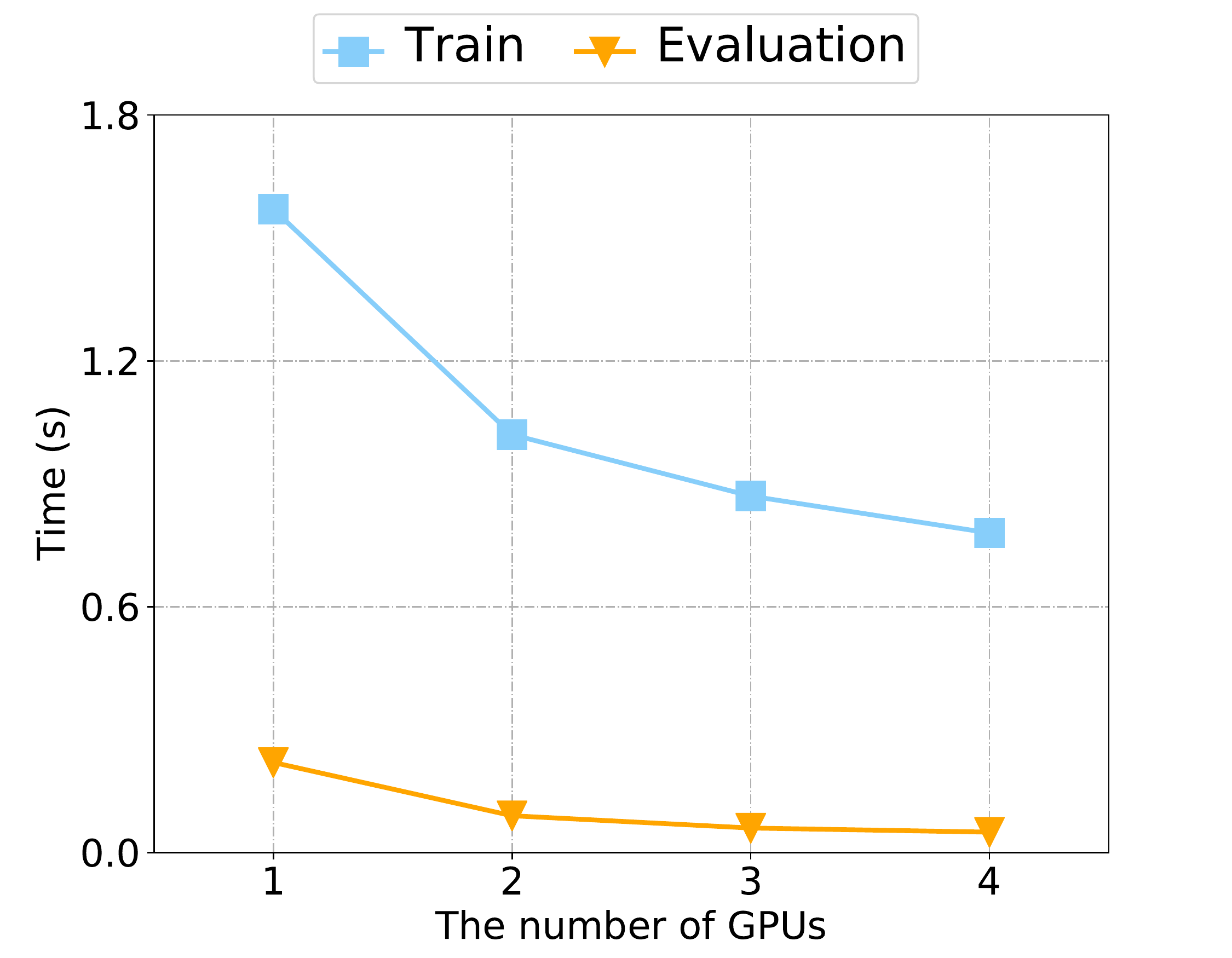}
    \captionsetup{font={small}}
    \caption{Efficiency comparison with different numbers of GPUs.}
    \label{fig:eff}
\end{figure}

We empirically analyze the training and evaluation costs of the BPR model~\cite{BPR} on MovieLens-1M~\cite{harper2015movielens} dataset with different number of GPUs.
As shown in Figure~\ref{fig:eff}, we can observe that as the number of GPUs increases, the training and evaluation time gradually decreases, showing the acceleration effect of distributed computing.

\subsubsection{Mixed precision training.}

\begin{table*}[h]
  \centering
  \small
  \caption{Performance comparisons of the CFKG and NNCF \emph{without}/\emph{with} mixed precision training~(MPT).}\label{tab:mptcom}
  \begin{tabular}{c|c|cccccc}
    \toprule
    \hline
    \multirow{2}{*}{\textbf{Model}} &
    \multirow{2}{*}{\textbf{MPT}}   &
    \multicolumn{6}{c}{\textbf{Metric}}                                                                                            \\
                                    &            & Recall@10 & MRR@10 & NDCG@10 & Hit@10 & Precision@10 & Training time~(sec/epoch) \\
    \hline
    \multirow{2}{*}{\textbf{CFKG}}  & \emph{w/o} & 0.1708    & 0.4061 & 0.2333  & 0.7108 & 0.1755       & \ \ 3.90                     \\
                                    & \emph{w/}   & 0.1708    & 0.4061 & 0.2333  & 0.7108 & 0.1755       & \ \ 3.41                     \\
    \hline
    \multirow{2}{*}{\textbf{NNCF}}  & \emph{w/o} & 0.1753    & 0.4293 & 0.2446  & 0.7232 & 0.1829       & 10.30                    \\
                                    & \emph{w/}  & 0.1770    & 0.4354 & 0.2471  & 0.7268 & 0.1836       & \ \ 5.47                     \\
    \bottomrule
  \end{tabular}
\end{table*}

\begin{table*}[t]
  \centering
  \small
  \caption{Performance comparisons of hyper-tuning strategies for two selected models.}\label{tab:hypercom}
  \begin{tabular}{c|c|c|ccccc}
    \toprule
    \hline
    \multirow{2}{*}{\textbf{Model}}&
    \multirow{2}{*}{\textbf{Tuning Strategy}}&
    \multirow{2}{*}{\textbf{Searching Time~(seconds)}}&
    \multicolumn{5}{c}{\textbf{Metric}}\\
     & & &Recall@10&MRR@10&NDCG@10&Hit@10&Precision@10 \\
     \hline
    \multirow{3}{*}{\textbf{LightGCN}} & \textbf{Grid Search} & \ \ 21,875.00 & 0.1853  & 0.4321 & 0.2487  & 0.7350 & 0.1833 \\
    & \textbf{Random Search} & \ \ 15,611.00 & 0.1851 & 0.4331 & 0.2498  & 0.7386 & 0.1848 \\
    & \textbf{Bayesian Hyper-opt} & \ \ 16,073.00 & 0.1823 & 0.4312 & 0.2476  & 0.7316 & 0.1835 \\ 
    \hline
    \multirow{3}{*}{\textbf{KGAT}} & \textbf{Grid Search} & 198,942.00  & 0.1867 & 0.4333 & 0.2521  & 0.7386 & 0.1866 \\
    & \textbf{Random Search} & 142,813.00  & 0.1816 & 0.4201 & 0.2456  & 0.7285 & 0.1841 \\
    & \textbf{Bayesian Hyper-opt} & 150,900.00 & 0.1845 & 0.4267 & 0.2498  & 0.7356 & 0.1863 \\
    \bottomrule
\end{tabular}
\end{table*}

Mixed precision training is an optimization technique for model training. It combines half-precision (FP16) and single-precision (FP32) with two major merits. For training efficiency, 
FP16 throughput in recent GPUs using Tensor Core technology is 2 to 8 times higher
than that for FP32~\cite{micikevicius2017mixed}, leading to a more efficient training framework; for memory cost,   the storage size required for FP16 is the half of FP32, which greatly reduces the memory cost.
To implement the mixed precision training, we use \textsf{torch.autocast} during the loss optimization to convert weights and gradients from FP32 to FP16. We also employ \textsf{torch.cuda.GradScaler} in order to scale loss and avoid overflow.

As shown in Table~\ref{tab:mptcom}, we conduct experiments to verify the performance of CFKG~\cite{ai2018learning} and NNCF~\cite{bai2017neural} \emph{with} and \emph{without} mixed precision training on MovieLens-1M dataset. 
We can observe that mixed precision training can largely reduce the training time, while retaining the recommendation performance. With mixed precision training, the speed is increased by 14.4\% on CFKG and 88.3\% on NNCF.

\subsubsection{Hyper-parameter Searching.} 
Neural recommendation models often rely on the hyper-parameter tuning to find the optimal parameter configurations. 
Compared with the previous version of RecBole, which mainly supports \textit{grid search}, this update provides users with more types of hyper-parameter tuning strategies, \ie \textit{Random Search~\cite{2012Random}} and \textit{Bayesian Hyper-opt~\cite{2012Practical}}, and we provide usage examples of tuning methods in the appendix.
For detailed analysis, we conduct experiments to compare the performance of three tuning strategies. Table~\ref{tab:hypercom} summarizes the performance of LightGCN~\cite{he2020lightgcn} and KGAT~\cite{wang2019kgat} tuned with different strategies on the MovieLens-1M dataset. 
We can observe that both \textit{Random Search} and \textit{Bayesian Hyper-opt} can generally find a good setting with less searching time than \textit{Grid Search} that enumerates all candidate options.
Furthermore, by introducing the unified framework \textsf{Ray}~\cite{Ray-tune} for scaling AI and Python applications, we now support parallel tuning with multiple GPUs for more efficient parameter search, and users can easily switch between parallel and serial parameter tuning via different configurations. The detailed examples of the two hyper-tuning tools are also presented in the appendix.

\subsection{Reproducible Configurations}

For a benchmarking recommendation library, the reproducibility of models is the core concern of users.
To fairly compare various methods, there are three key aspects to consider, including benchmarking datasets, unified tools and optimal configurations. 
The previous version of RecBole mainly focuses on the first two aspects, and this update further expands the number of processed benchmarking datasets from 28 to 41 and releases the recommended hyper-parameter settings for model reproducibility.

\subsubsection{Extended Benchmarking Datasets}
In RecBole, we use the atomic files to format the public datasets in a unified way, which is convenient to maintain and manipulate. To further extend the dataset availability in RecBole, we add 13 new datasets  on the basis of 28 existing ones (Table~\ref{tb:new-dataset}). For details of our datasets and scripts, please refer to our dataset repository at the link: \url{https://github.com/RUCAIBox/RecSysDatasets}. As shown in Table~\ref{tb:new-dataset}, these new datasets have the following features:

$\bullet$ \textbf{Multi-domain datasets}. The newly added datasets for recommendation cover the categories of e-commerce, movies, restaurants and books, which enriches the research scenarios of our library. 

$\bullet$ \textbf{Multi-version datasets}. For the same recommendation dataset, there may be multiple versions with different sizes and time periods.
To clarify multi-version datasets, we collect different versions of two commonly used datasets, \ie Yelp~\cite{Yelp} datasets published in 2018, 2020, 2021 and 2022, and Amazon review datasets~\cite{Amazon} of 24 sub-categories in 2014 and 29 sub-categories in 2018.
With this improvement, researchers can easily select multiple versions of benchmarking datasets, from the classic to the latest one.

$\bullet$ \textbf{Rich-context datasets}. 
The context information of the users and items (\eg item attributes) is important for content or context based recommendation tasks. 
Most of the updated datasets contain additional feature information for users or items, which can well support the research on  context-based recommendation.

\begin{table}[H]
\centering
\caption{Newly collected datasets in RecBole.}\label{tb:new-dataset}
\begin{tabular}{@{}crrr@{}}
\toprule
Dataset           & \#Users     & \#Items    & \#Interactions \\ \midrule
Amazon (2018)     & -  & -    & -    \\
Alibaba-iFashion  & 3,569,112  & 4,463,302 & 191,394,393   \\
AliEC             & 491,647    & 240,130   & 1,366,056     \\
BeerAdvocate      & 33,388     & 66,055    & 1,586,614     \\
Behance           & 63,497     & 178,788   & 1,000,000     \\
DianPing          & 542,706    & 243,247   & 4,422,473     \\
EndoMondo         & 1,104      & 253,020   & 253,020       \\
Food              & 226,570    & 231,637   & 1,132,367     \\
GoodReads         & 876,145    & 2,360,650 & 228,648,342   \\
KGRec-music       & 5,199      & 8,640     & 751,531       \\
KGRec-sound       & 20,000     & 21,533    & 2,117,698     \\
ModCloth          & 47,958     & 1,378     & 82,790        \\
RateBeer          & 29,265     & 110,369   & 2,924,163     \\
RentTheRunway     & 105,571    & 5,850     & 192,544       \\
Twitch            & 15,524,309 & 6,161,666 & 474,676,929   \\
Yelp (4 versions) & 5,556,436  & 539,254   & 28,908,240    \\ \bottomrule
\end{tabular}
\end{table}

\subsubsection{Recommended Hyper-Parameter Settings}
As a unified recommendation library, RecBole provides researchers with a standardized framework to reproduce baselines and design new models.
However, as shown in the feedback of our users, it is still time-consuming  to search for the optimal hyper parameters that lead to appropriate model results.
To facilitate users to reproduce model results, we have tried our best to make the implementations adhere to the paper and source code~(if any), and make a thorough parameter search for performance tuning.   
In this update, we  release the hyper-parameter tuning range and recommended configurations for each model on three benchmarking datasets respectively.
According to the hyper-parameter settings in \url{https://recbole.io/hyperparameters}, researchers can directly  get the model results on benchmarking datasets, which largely eases the model tuning. 
Next, we introduce our efforts in  reproducing the benchmarking results. 

$\bullet$ \textbf{Representative datasets}. For each of four types of recommendation tasks in RecBole, we provide three benchmarking datasets respectively. For general and sequential recommendation, we adopt classic movie recommendation dataset MovieLens-1M~\cite{harper2015movielens}, the latest Yelp dataset~\cite{Yelp} in 2022 and Amazon Books dataset~\cite{Amazon} in 2018, considering the popularity, domains and scales. For context-aware recommendation, we also adopt MovieLens-1M~\cite{harper2015movielens} since it contains both temporal and attribute information for movies. Besides, Criteo~\cite{Criteo} and Avazu~\cite{Avazu} are two frequently used datasets for click-through rate predictions, which are used as the benchmarking datasets for context-aware recommendation. For knowledge-aware recommendation, we use three datasets in KB4Rec~\cite{zhao2019kb4rec} linked to Freebase~\cite{Freebase} as our benchmarking datasets, which are widely adopted by academia on knowledge-aware recommendation~\cite{wang2019kgat,KERL,ZhouImproving20}.
To ensure the data quality, we apply 10-core filtering for users and items to remove inactive users and unpopular items. We also employ 5-core filtering for nodes and relations in knowledge-aware models. As for data  splitting, we utilize the \textit{leave-one-out} strategy in temporal order for sequential models, while we split data into train/validation/test sets with a ratio of 8:1:1~\cite{zhao2019revisiting} for the other three types of models.

$\bullet$ \textbf{Comprehensive models}. For the four types of recommendation tasks in RecBole, we release the recommended hyper-parameter tuning settings on three benchmarking datasets for each type, including general, sequential, context-aware and knowledge-aware recommendation.
And meantime, our team welcomes community contributors to work together for maintaining and updating the benchmarking configurations~(especially the original authors). In the near future, we consider publishing the optimal hyper-parameter settings of the eight sub-packages in RecBole 2.0~\cite{zhao2022recbole} to facilitate follow-up research. 

$\bullet$ \textbf{Fine-grained hyper-parameter tuning}. In order to obtain the optimal (or nearly optimal) performance for a given recommendation model, we utilize the grid search strategy for parameter tuning on each dataset. As for the search range of hyper-parameters, we follow the settings of original papers and fine-tune specific parameters according to their training methods. To ensure a fair comparison, common hyper-parameters such as the learning rates and regularization values are tuned in the same range for all the models on the same dataset. 
Meanwhile, the hyper-parameters of the same model architecture remain the same, such as the number of convolution layers in GNN-based models and the size of attention heads in Transformer-based models. 

$\bullet$ \textbf{Reproducible configurations}. To ensure the reproducibility, we release the configuration files of datasets and models, the hyper-parameter tuning range and recommended model  parameters. For hardware environment, our experiments are all conducted on an Nvidia GPU with 12 GB memory, with CUDA 11.4. Based on these public configurations, researchers can reproduce the benchmarking results of different models conveniently. 

Note that, in this update,  we do not directly release the benchmarking results, but instead report  the hyper-parameter search range and the recommended parameter configuration.

\subsection{User-Friendly Documentation}

As a user-friendly recommendation library, RecBole maintains the code repository, reference papers, project websites, reproducible configurations and API documentation. In this update, 
we largely update the website and documents to improve the user experience. 

$\bullet$ \textbf{Comprehensive website}. As a major change, we update the website~(\url{https://recbole.io}) by officially including RecBole 2.0. Specially, we add a a web page for all the sub-packages in RecBole 2.0, so as to access more than 130 models included in the two versions. Also, we add a navigational  table for 11 related repositories of RecBole on GitHub for a quick  reference, 
so that users can quickly find the needed resource in RecBole.

$\bullet$ \textbf{Detailed tutorials}. To improve user experiences, 
our updated handbook includes detailed illustrations for both basic usage and new features. As for model training, we add instructions of customized training strategy to facilitate user-defined trainer. Meanwhile, we add usage examples to demonstrate the multi-GPU training, including single GPU, single machine multi-GPU and multi-machine multi-GPU. For a quick reference of the library usage, we also provide detailed running steps for each type of models.

$\bullet$ \textbf{Open community}. Since our release in 2020, RecBole has received extensive attention from open-source contributors on GitHub, with nearly 2300 stars and 425 forks. Our team has been following up the  research advances,  responding timely to the issues of users on GitHub~(472 issues were solved until Nov 28, 2022). We warmly welcome any contribution to RecBole in code contribution, documentation translation, and use experience sharing.  Our team is looking forward to your contributions to RecBole and feel free to contact us by email recbole(@)outlook.com. 














\section{DISCUSSION AND CONCLUSIONS}

In this report, we upgrade RecBole, a popular open-source recommendation library from version 1.0.1 to 1.1.1. Our update highlights can be featured in four aspects, \ie more flexible data processing, more efficient training and tuning, more reproducible configurations and more user-friendly documentation. RecBole is expected to help researchers easily reproduce baselines as well as develop new algorithms, serving as a comprehensive open source benchmarking library for recommendation. In the future, we will continuously update RecBole to make it more \textit{flexible}, \textit{up-to-date} and \textit{user-friendly}.

\paratitle{Acknowledgement}.
We sincerely thank the support from all users and previous developing members of RecBole.

\newpage
\bibliographystyle{ACM-Reference-Format}
\bibliography{acmart}

\newpage
\appendix
\section{APPENDIX}

\begin{figure}[hb]
    \centering
\begin{lstlisting}[language=python]
from recbole.quick_start import run_recboles
import torch.multiprocessing as mp
args = dict(
    # General parameters in original RecBole
    model = 'BPR',
    dataset = 'ml-100k',
    config_file_list = ['test.yaml'],

    # Parameters for distributed training
    ip = 'IP address of the master node',
    port = 'Port number of the master node', 
    world_size = 'Number of ranks in the global', 
    nproc = 'Number of processes on current node', 
    group_offset = 'Offset of the global rank'
)
mp.spawn(
    run_recboles,
    args=(
        args.model,
        args.dataset,
        config_file_list,
        args.ip,
        args.port,
        args.world_size,
        args.nproc,
        args.group_offset,
    ),
    nprocs=args.nproc,
)
\end{lstlisting}
    \caption{Usage example of Distributed Training}
    \label{fig:ddp}
\end{figure}

\begin{figure}[hb]
\begin{lstlisting}[language=python]
python run_hyper.py --tool=Ray
python run_hyper.py --tool=Hyperopt
\end{lstlisting}
    \caption{Usage examples of two hyper-tuning tools.}\label{fig:hyper-example}
\end{figure}

\begin{figure}[hb]
    \subfigure[Grid Search]{%
\begin{lstlisting}[language=python]
from recbole.trainer import HyperTuning

hp = HyperTuning(
    objective_function,
    algo="exhaustive", # tuning strategy
    early_stop=10, # early stop steps for tuning
    max_evals=100, # maximum of searches, not used here
    params_file=args.params_file,
    fixed_config_file_list=config_file_list,
    display_file=args.display_file, # display output path
)



\end{lstlisting}
        \label{fig:grid}
    }
    \subfigure[Random Search]{%
\begin{lstlisting}[language=python]
from recbole.trainer import HyperTuning

hp = HyperTuning(
    objective_function,
    algo="random", # tuning strategy
    early_stop=10, # early stop steps for tuning
    max_evals=100, # maximum of searches
    params_file=args.params_file,
    fixed_config_file_list=config_file_list,
    display_file=args.display_file, # display output path
)



\end{lstlisting}
        \label{fig:random}
    }
    \subfigure[Bayesian Hyper-opt]{%
\begin{lstlisting}[language=python]
from recbole.trainer import HyperTuning

hp = HyperTuning(
    objective_function,
    algo="bayes", # tuning strategy
    early_stop=10, # early stop steps for tuning
    max_evals=100, # maximum of searches
    params_file=args.params_file,
    fixed_config_file_list=config_file_list,
    display_file=args.display_file, # display output path
)



\end{lstlisting}
        \label{fig:bayes}
    }
    \caption{Usage examples of hyper-tuning strategies.}
    \label{fig:hyper-opt}
\end{figure}

\end{document}